\documentclass[USenglish,twocolumn,final]{article}

\usepackage[utf8]{inputenc}
\usepackage[big]{dgruyter}
\usepackage{microtype}

  \usepackage{graphicx}
  \usepackage{mathtools}
  \usepackage{bm}
  \usepackage{upgreek}
  \usepackage{cancel}
  \usepackage{color}
  \usepackage{siunitx}
  \usepackage{gensymb}
	\usepackage{acronym}
  \usepackage{textcomp}
  \usepackage[normalem]{ulem}
  
  \usepackage{soul}
  \sethlcolor{black}
  \usepackage{cite}

\definecolor{gray}{gray}{0.6}
\definecolor{cblue}{rgb}{.7,.7,.9}

\newacro{FIB}[FIB]{Focus Ion Beam}
\newacro{HHG}[HHG]{Higher Harmonic Generation}
\newacro{THG}[THG]{Third Harmonic Generation}
\newacro{FWM}[FWM]{Four Wave Mixing}
\newacro{NSOM}[NSOM]{Near-field Scanning Optical Microscope}

\begin{document}


  \articletype{Review article}

  \author[1,2]{Kirill Koshelev}
  \author[3]{Gael Favraud}
  \author[2]{Andrey Bogdanov}
  \author[1,2]{Yuri Kivshar}
  \author*[3]{Andrea Fratalocchi}
  \affil[1]{Nonlinear Physics Center, Australian National University, Canberra ACT 2601, Australia}
    \affil[2]{ITMO University, St. Petersburg 197101, Russia}
      \affil[3]{PRIMALIGHT,
  	Faculty of Electrical Engineering, Applied Mathematics and Computational Sciences,
  	King Abdullah University of Science and Technology (KAUST), Thuwal 23955-6900, Saudi Arabia}
  \title{Nonradiating Photonics with Resonant Dielectric Nanostructures}
  \runningtitle{Nonradiating photonics}
\abstract{
Nonradiating sources of energy have traditionally been studied in quantum mechanics and astrophysics, while receiving a very little attention in the photonics community. This situation has changed recently due to a number of pioneering theoretical studies and remarkable experimental demonstrations of the exotic states of light in dielectric resonant photonic structures and metasurfaces, with the possibility to localize efficiently the electromagnetic fields of high intensities within small volumes of matter. These recent advances underpin novel concepts in nanophotonics, and provide a promising pathway to overcome  the problem of losses usually associated with metals and plasmonic materials for the efficient control of the light-matter interaction at the nanoscale. This review paper provides the general background and several snapshots of the recent results in this young yet prominent research field, focusing on two types of nonradiating states of light that both have been recently at the center of many studies in all-dielectric resonant meta-optics and metasurfaces: optical {\em anapoles} and photonic {\em bound states in the continuum}. We discuss a brief history of these states in optics, their underlying physics and manifestations, and also emphasize their differences and similarities. We also review some applications of such novel photonic states in both linear and nonlinear optics for the nanoscale field enhancement, a design of novel dielectric structures with high-$Q$ resonances, nonlinear wave mixing and enhanced harmonic generation, as well as advanced concepts for lasing and optical neural networks.
}
  \journalname{Nanophotonics}
  \startpage{1}
  \aop

\maketitle

\section*{Introduction}

Nonradiating sources are defined as particular electromagnetic states that do not emit energy in the far-field region. They originate from oscillating charge-current configurations that do not generate electromagnetic fields outside a localized volume. Traditionally, nonradiating sources did not attract much attention in photonics, because it was believed that such states do not permit the energy propagation away from the source. On the contrary, the exotic states of matter attracted a lot of attention in the field of quantum mechanics in the context of radiationless motion, extended electron models, and radiation reactions, also providing possible scenarios for observing a dynamic form of the well-known Aharonov-Bohm effect \cite{PhysRev.74.1789,PhysRev.135.B281,Pearle1978,doi:10.1119/1.14084,Davidson2007,Nemkov-2017-PRB}.

Recently, research on nonradiating electromagnetic states has become the subject of intense studies in classical electrodynamics and, in particular, at optical wavelengths. Such an interest has been boosted by many studies of light-matter interactions in all-dielectric resonant nanoparticle systems supporting the electric and magnetic multipolar Mie-type resonances, in efforts to overcome the low-quality factors of their resonant modes (see, e.g., Ref.~\cite{Rybin-2017-PRL} as an example, and many references below). The results obtained in this field very recently revealed that this platform allows creating nonradiating states of light in high-index dielectric structures with the subwavelength confinement properties similar to metallic and nanoplasmonic structures~\cite{Maier:2007:10.1007/0-387-37825-1}, but with much weaker or almost negligible losses~\cite{Kivshar-2018-NSR}.

In the context of nonradiating sources, the theory developed during the 90-s demonstrated that different types of nonradiating states could be generated by combining classical electric and magnetic dipole modes with electromagnetic toroidal multipoles, in the configurations where the angular distribution of the energy density has the same and opposite parities \cite{PhysRevD.35.3496,1998PPN....29..366A}. The simplest of these configurations is created by {\it a toroidal dipole mode} oscillating out of phase with respect to a single electric dipole.  It was noted that such a particular interfering state forms a simple but nontrivial nonradiating point source, later recognized as an elementary dynamic anapole state, or simply {\it anapole}~\cite{Papasimakis-2016-NM}.

Despite these results were available for some time in the literature, it took more than thirty years to demonstrate experimentally the existence of these exotic states, first at microwave frequencies~\cite{fedotov2013resonant} and later, in a single-particle geometry, at optical frequencies~\cite{Miroshnichenko-2015-NC,PhysRevB.91.035116}, followed by an extensive theoretical work generalizing these findings to different materials and settings~\cite{basharin2015dielectric,Liu:15,PhysRevB.94.205433,PhysRevB.94.205434}, also opening a direct link to optical invisibility~\cite{Feng2007,doi:10.1002/adma.201202624,Wang:14,Rybin2015}.

Anapoles are not the only radiationless states that can be observed in meta-optics and nanoscale photonics. Theoretical and experimental studies revealed that the complex interaction of multimode and multipolar resonances in Mie-resonant dielectric structures can also support a different class of nonradiating electromagnetic states characterized by no energy leakage and spatially localized profiles. These states are characterized by the resonant frequencies lying in the continuum spectrum of radiating modes of the structure, and they are known as {\it bound states in the continuum} (BIC) from earlier days of quantum mechanics~\cite{Neumann-1929-PZ}. In spite of original BICs were never observed experimentally, their beautiful properties attracted wide attention in atomic physics~\cite{fonda1963bound}, hydrodynamics and acoustics~\cite{ursell1951trapping, cumpsty1971excitation}, and, in recent years, in photonics. Importantly, being substantially different from anapoles, BICs share similar properties in terms of the electromagnetic energy trapping and effective localization of light in small volumes of matter. 

In nanophotonics, BICs are associated with "dark optical states" which can be supported by a special class of all-dielectric nanoscale structures. While the first BICs were discussed in optics only in 2008~\cite{Marinica-2008-PRL, bulgakov2008bound}, they were followed by the experimental demonstration in coupled waveguides only in 2011~\cite{Plotnik-2011-PRL}. To date, the properties of optical BICs are being studied extensively for different designs and geometries, including filtering, lasing, and sensing applications~\cite{foley2014symmetry, Kodigala-2017-N, liu2017optical}. Practical implementation of BICs for nonlinear optics, twisted light, and light-mater interaction is under active study~\cite{bulgakov2014robust, bulgakov2017bound, koshelev2018strong}. 

We should notice here that a rigorously defined nonradiating state of light remains a pure mathematical concept unless such a localized state is coupled to an external radiation being transformed into {\em a leaky mode}. Such quasi-nonradiating states are based on the same physics of the radiation compensation and destructive interference, but they become practically very useful for enhancing the light-mater interaction. In experiments, both quasi-anapoles and quasi-BIC are observed because of a luck of complete compensation due to material absorption, finite-size samples, structural disorder, and surface scattering~\cite{Miroshnichenko-2015-NC, Sadrieva-2017-ACSP, belyakov2018experimental}. Thus, in practice, an anapole is recognized as a dynamic current distribution with the compensation or absence of the leading multipole orders, while a BIC transforms into a quasi-BIC or supercavity mode~\cite{Rybin-2017-N}, that possesses high yet finite quality factor ($Q$ factor) and large but limited field enhancement.

Accessing quasi-nonradiating photonic states in all-dielectric resonant nanostructures that support localization of light and large $Q$ factors opened up many applications in photonics. Despite this research is only a few years old, it already counts a large number of theoretical predictions and experimental demonstrations in many different areas, ranging from advanced material engineering and functional metasurfaces to nonlinear effects sustained by complex multimode mixing at different frequencies. The aim of this review is to highlight some of the recent results in this new recently emerged research area, providing a up-to-date overview of the recent applications and possible future perspectives.

\section*{Nonradiating states of light}

\subsection*{Anapoles}

The term {\it anapole} means ``without poles'', and it was originally proposed in the late 50's by Jacob Zeldovich~\cite{ZelDovich-1958-SJETP} to model elementary particles which do not interact with the electromagnetic fields. Anapole states have been investigated in astrophysics, as they could explain the origin and composition of dark matter in the universe within the classical physics~\cite{Ho-2013-PLB}.

The concept of anapoles was generalized to electrodynamics in the theoretical studies of Dubovik, Afanasiev and co-workers, who suggested that such states could be created by a suitable combination of toroidal dipoles mixed with electric and magnetic multipoles, leading to the formation of different possible nonradiating states~\cite{Dubovik-1990-PR,Afanasiev-1995-JPA}.

The first structure that experimentally demonstrated an anapole mode in electrodynamics was reported for a metallic metamaterial made of thin plates characterized by specific perforations to generate toroidal modes inside the structure \cite{fedotov2013resonant}.

The first experimental demonstration of the anapole at optical wavelengths was reported in 2015 for a dielectric silicon nanodisk~\cite{Miroshnichenko-2015-NC}. In this work,  the authors employed an analytical multipole expansion using spherical multipoles to describe the field outside of the nanodisk, and the Cartesian multipoles for describing the currents inside, and they confirmed the existence of the special condition for an anapole state in spherically symmetric structures made of sufficiently high refractive index materials (typically $n\ge 3$).

\begin{figure}[t]
  \centering
  \includegraphics[width=\columnwidth]{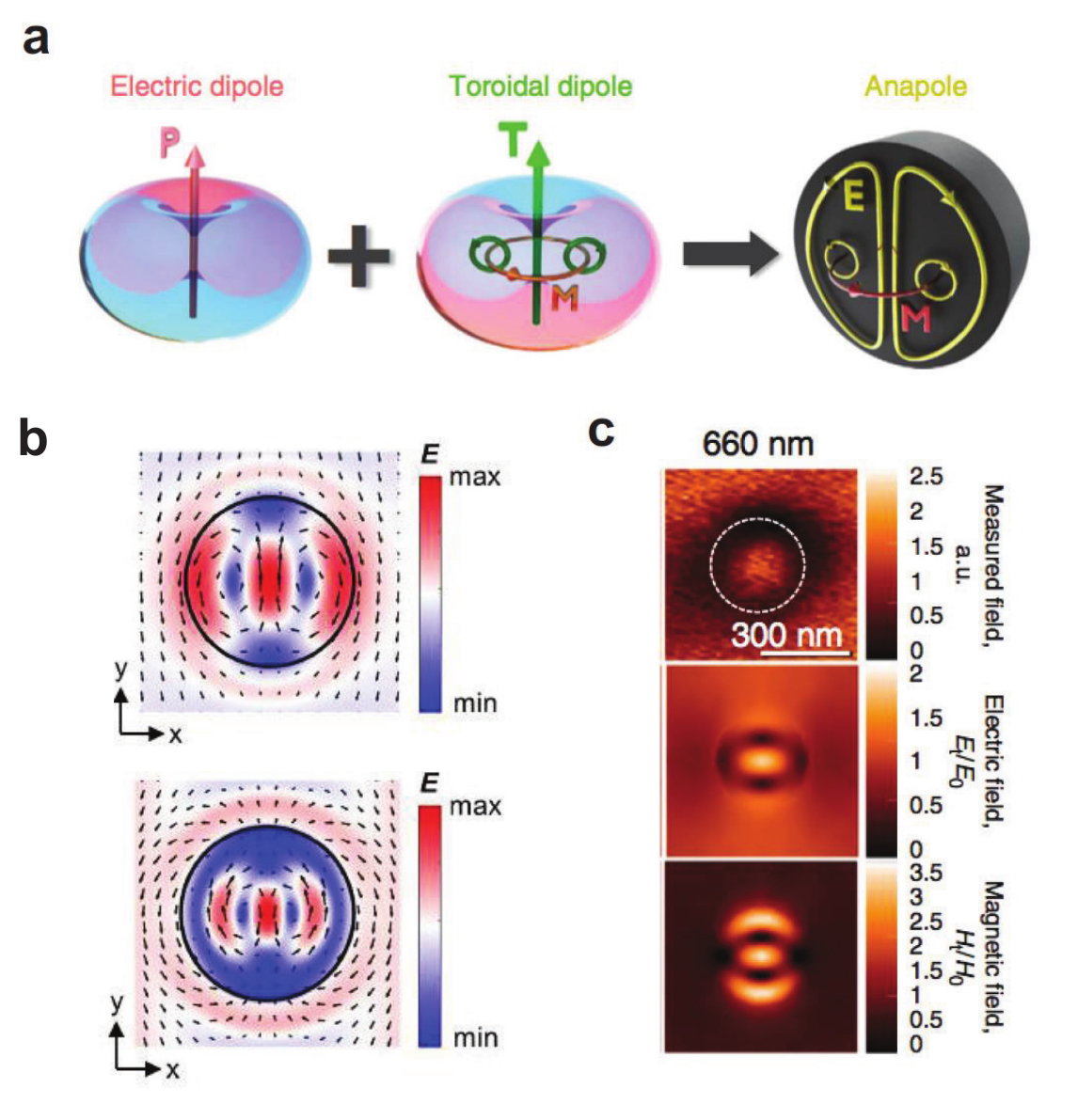}
  \caption{
      \label{fig1}
\textbf{Physics and observation of the anapole state.} \textbf{(a)} Conceptual representation of the anapole state as a superposition of electric dipole and toroidal dipole modes. A special feature of these two dipoles is that their far-field distribution is identical. For a particular set of parameters, these two dipoles can completely cancel each other by destructive interference in the far-field region when oscillating out of phase, hence creating a nonradiating anapole.  \textbf{(b)} Numerical field structure of the two types of anapoles excited in a dielectric nanodisk. \textbf{(c)} Experimental SNOM measurements of the anapole state. The contour of the nanodisk is represented with a white dashed line. The middle and bottom images show the computed  transverse electric and magnetic near-field, respectively, 10~nm above the nanodisk. Adapted with permission from Ref.~\cite{Miroshnichenko-2015-NC}.
	}
\end{figure}

As this type of structure is not particularly suited for experimental observation under the plane wave illumination,
the authors generalized these results to cylindrically shaped nanodisks possessing a particular aspect ratio. They demonstrated that, by optimizing the height and radius of each nanodisk, it is indeed possible to suppress higher order multipoles and observe the direct interference of the toroidal and electric dipoles [see Figs. \ref{fig1}a,b] at a specific (anapole) frequency, which can be tuned in the visible and near infrared.

Theoretical results were confirmed through scanning near-field optical microscope (SNOM) experiments [see Fig.~\ref{fig1}c] with $\SI{50}{nm}$ thick silicon nanodisks of diameters varying from 200 to $\SI{400}{nm}$. These nanodisks manifested an invisible behavior in the far field at the frequency of the anapole revealed by a local minima in their scattering cross section.
SNOM experiments showed that as the impinging wavelength tends to the reported anapole wavelength $\lambda=\SI{640}{nm}$, a hot-spot appeared in the center of the nanodisk in very good agreement with theory and numerical simulations, providing the first experimental observation of the anapole state [see Fig. \ref{fig1}c].

After that experimental observation, the anapole states were demonstrated in dielectric nanoscale spheres of the radius $\SI{100}{nm}$ placed at the focal point of a tightly focused polarized beam, realizing a standing wave by superposition of two counter-propagating wave at a wavelength of $\lambda=\SI{464}{nm}$~\cite{Wei-2016-O}, and
 in core-shell structures such as nanowires~\cite{Liu:15}.
 
A large portion of the theoretical studies of the anapole states relies traditionally on complex multipole expansions of the Mie theory~\cite{Papasimakis-2016-NM}. This makes the generalization to geometries lacking cylindrical or spherical symmetries quite challenging. For this reason, several groups of authors started to employ some other approaches. In particular, Nemkov and co-authors~\cite{Nemkov-2017-PRB} derived a formulation of the anapole states based on the vector potential, which allows to rewrite nonradiating sources as a distribution of elementary anapoles. This method also provides a very intuitive approach to understand and analyze nonradiating sources in general.

\begin{figure*}[t]
  \centering
  \includegraphics[width=16cm]{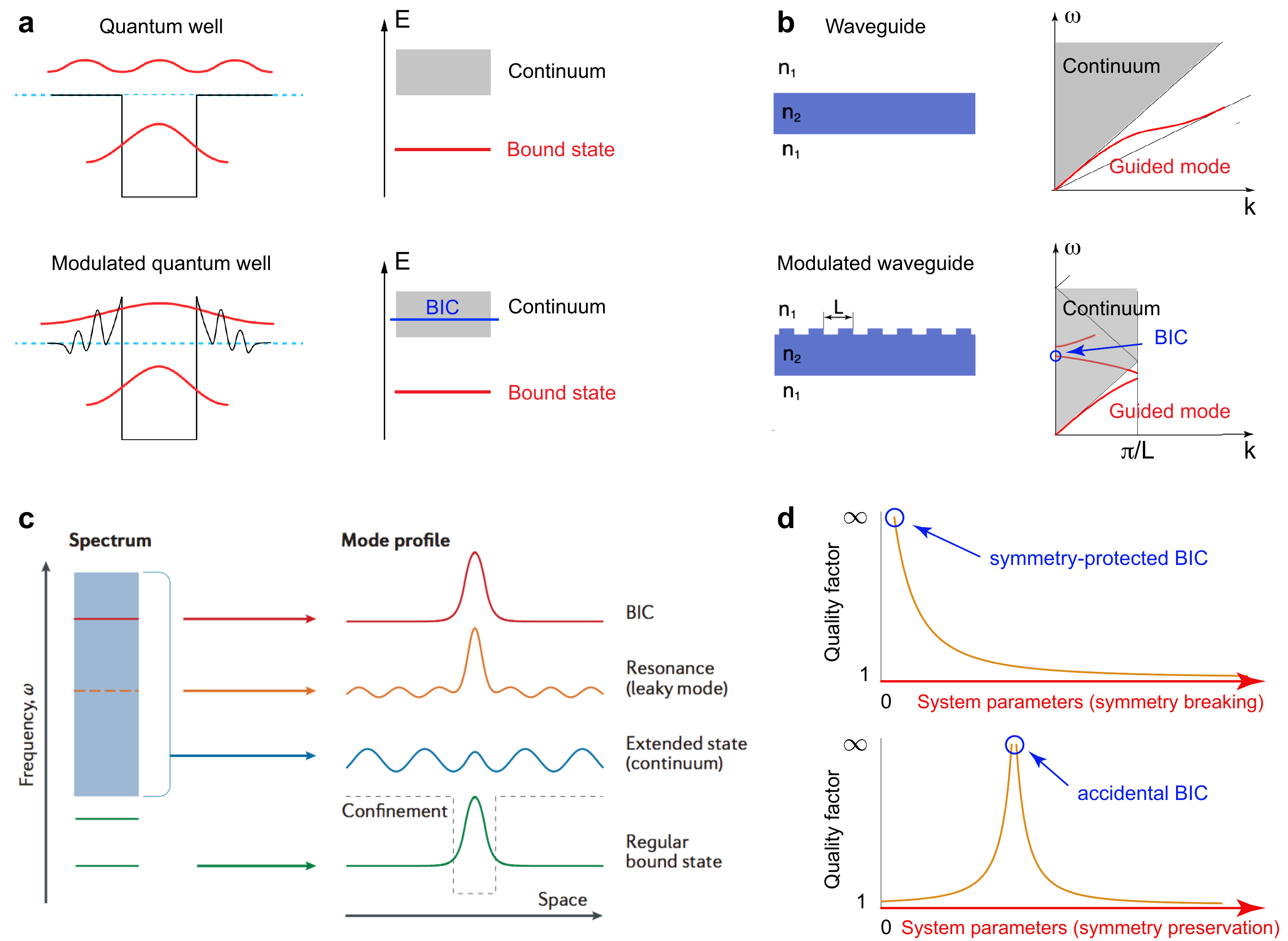}
  \caption{
      \label{fig2}
       \textbf{Concept of bound states in the continuum.}
       (a) Schematic illustration of origin of BICs in quantum mechanical systems. Modulation of the potential of a conventional quantum well (top panel) with spatially unbound oscillations (lower panel) allows for localization of a bound state within the continuous spectral range (BIC, shown in blue). 
       (b) Schematic illustration of origin of BICs in photonic systems. Here, emergence of a BIC (shown in blue) within the continuous spectral range is due to modulation of the dielectric function of a regular waveguide (top panel) with a periodic perturbation (lower panel). 
       (c) Different families of states sustained in an open photonic system, and their position in the spectrum: regular bound states, extended radiation continuum modes, leaky resonances and BICs. 
       (d) Different mechanisms of BIC formation: due to symmetry protection (upper panel) and parameter tuning (lower panel). A symmetry-protected BIC transforms into a leaky resonance with a finite $Q$ factor when the symmetry is broken. An accidental BIC emerges via continuous tuning of the system parameters which preserve the structural symmetries.
      Part (c) is adapted with permission from~\cite{Hsu-2016-NRM}.
	}
\end{figure*}

Another approach targeted to build anapoles from general geometries was proposed by Totero and co-workers~\cite{Totero-2017-N}. The main point in that work was to provide a fast and general approach for the computation of electromagnetic modes in arbitrary-shaped open optical structures. This form a challenging problem because electromagnetic modes defined from radiating boundary condition do not form a Hermitian eigenvalue problem with complete, orthogonal, and easy to calculate modes such as, e.g., the modes in quantum mechanics. Despite new approaches were developed recently allowing for analytic mode normalization for non-Hermitian photonic systems~\cite{lalanne2018light}, their computation difficulty is still time consuming. Even in the context of the Mie modes, known algorithms tend to be unstable in the computation of high order modes due to the inherent complexity of the special functions describing these resonances.

Being inspired by the methods used to model open quantum systems, the authors of~\cite{Totero-2017-N} proposed a method based on a Fano-Feshbach projection scheme. It consists in describing the dynamics of Maxwell equations in terms of orthogonal and complete eigenmodes of an internal domain, interacting with scattering waves of an external domain. The main advantage of this approach is to rely on an orthogonal basis of well-behaved eigenmodes of ideal cavities, describing the time dynamics of the system with a simple set of exact coupled-mode equations. Under such description, an anapole state can be generated from the destructive interference between internal and external modes interacting on the surface separating internal and external domains. The authors applied this approach to anapoles in nanodisks, showing that this state can be interpreted from a complex interaction of two almost overlapping internal resonances. With this approach, the authors  described theoretically the existence of higher-order anapole modes [see Fig.~\ref{fig1}b], which could not be represented in terms of the fundamental anapole mode, but they originate from a complex superposition of several internal resonances of the structure.

As mentioned above, interference of the electric and toroidal dipole moments results in a very peculiar, low-radiating optical state associated with the concept of the nonradiating optical anapole. The physics of multimode interferences and multipolar interplay in nanostructures is expanding rapidly, and the intriguing example of the optical anapole is only one of examples of such states. The recently emerged field of {\em anapole electrodynamics} has been reviewed recently in Ref.~\cite{AOM_anapole} together with its relevance to multipolar nanophotonics and manifestations in nonlinear optics, active photonics, and metamaterials.

\subsection*{Bound states in the continuum}
Another family of nonradiating states supported by all-dielectric nanostructures are so-called bound states in the continuum (BICs).
BICs were originally introduced in the context of the Schr\"odinger equation in 1929 by von Neumann and Wigner~\cite{Neumann-1929-PZ}.
These states correspond to peculiar resonances that are localized within the radiation continuum spectrum, while having no energy decay. 
Due to the universal nature of the Schr\"odinger equation, BICs represent an ubiquitous phenomena pertaining to all domains of wave physics ranging from microwave to optics, acoustics, and water waves~\cite{Hsu-2016-NRM}, in which they were sometimes termed as "trapped modes","embedded eigenvalues" or "dark states"~\cite{Gentry:14,Hodaei:16}.\\

Figure~\ref{fig2}a illustrates the conceptual origin of BICs in quantum mechanics. The spectrum of a conventional quantum well consists of two families of modes, which are clearly separated in energies: delocalized states existing within the continuum of propagating solutions and bound states which form a discrete set of modes. Exotic bound in the continuum states, which are spatially localized despite their energies lie in the continuous part of spectrum, can be achieved by the specific modulation of the potential with unbound oscillations. The analogy can be drawn in photonics, where a BIC can be realized due to periodic patterning of a plain waveguide, as shown in Fig.~\ref{fig2}b. Figure ~\ref{fig2}c demonstrates the frequency bands and the spatial profiles for each category of states of an open photonic structure: regular bound states, extended states, leaky resonances and BICs. 

Being completely decoupled from the incoming radiation, BICs can be excited only by local sources inside the structure. They possess an infinite quality factor ($Q$ factor) representing dark optical states. Figure~\ref{fig2}d shows schematically two main approaches proposed to realize a BIC in nanophotonics, where the first is based on the symmetry induced suppression of radiation while the second is the accidental decoupling from the radiation continuum due to the continuous tuning of system parameters. Here, we summarize the central points of the work by Hsu and others~\cite{Hsu-2016-NRM} which reviewed the properties of different classes of BICs in details.

The first approach relies on the presence of structural symmetries. A typical example of such symmetry-protected BIC is the waveguide with a cavity tuned to the frequency for which all modes of one symmetry are evanescent. Under this condition, the localized state of the cavity possessing that particular symmetry is prohibited to radiate into the waveguide. Then, if the mode frequency lies within the continuum spectrum of radiating modes of another symmetry, the BIC condition is fulfilled. Another important example of symmetry-protected BICs was demonstrated for periodic photonic structures, such as gratings, photonic crystal slabs, metasurfaces, or chain arrays. In the case the structure obeys the time-reversal and the in-plane inversion symmetry, and the mode frequency is below the diffraction limit, a symmetry-protected BIC can be realized in high-symmetry points of the first Brillouin zone, for example, at $\Gamma$ point. Interestingly, defects and perturbations breaking the structural symmetries allow coupling of BICs to the radiative modes, transforming them into leaky resonances with a finite $Q$ factor which depends on the coupling strength.

Another important category of BICs is accidental BICs originating through tuning of some system parameters. They can be divided in two major groups: Fabry-P\'erot BICs based on two identical resonances of two interacting cavities coupled through a single radiation channel, and the Friedrich-Wingten BICs representing the interference of two different resonances in the same cavity~\cite{Friedrich-1985-PRA}. Importantly, a Friedrich-Wingten BIC can be regarded as a limiting case of the Fabry-P\'erot BIC, in which the length of the radiation channel connecting the two cavities vanishes.  For periodic photonic structures, accidental BICs of the Friedrich-Wingten type can be realized, if the mode frequency is below the diffraction limit and the structure is invariant under the time-reversal, in-plane inversion, and up-down mirror symmetries. For this case, it is possible to achieve an accidental BIC not only at the edges of the Brillouin zone, but at an arbitrary point of the reciprocal space. Thus, a propagating BIC with a finite group velocity can be realized with the frequency lying in the continuous range of the free space.

The behavior of accidental BICs can be theoretically explained within the temporal coupled mode theory applied to a simple system of two resonances, where the amplitudes $\mathbf a=[a_1(t), a_2(t)]^T$ of the modes evolve in time as $\frac{d\mathbf a}{dt}=\mathcal H\mathbf a$ with the following Hamiltonian
\begin{equation}
\label{ham0}
\mathcal H=\begin{bmatrix}
\omega_1 &\kappa\\
\kappa &\omega_2
\end{bmatrix}-i\begin{bmatrix}
\gamma_1 &\sqrt{\gamma_1\gamma_2}e^{i\psi}\\
\sqrt{\gamma_1\gamma_2}e^{i\psi} &\gamma_2
\end{bmatrix}.
\end{equation}
Here, $\omega_i$, $\gamma_i$ are the resonant frequencies and the damping rates of the modes $i=1,2$, respectively, $\psi$ is the phase shift between the modes and $\kappa$ is the coupling factor. 
One of the eigenvalues becomes purely real with no decay, turning into a BIC, when the following conditions are attained
 \begin{align}
&(\omega_1-\omega_2)\kappa=e^{i\psi}\sqrt{\gamma_1\gamma_2}(\gamma_1-\gamma_2),\\
&\psi = \pi m,
\end{align}
where $m$ is an integer number. This conditions can be fulfulled through the tuning of the parameters of two coupled resonances in the case of a Friedrich-Wingten BIC, or in the case of a Fabry-P\'erot BIC when the two resonances are the identical $\omega_1=\omega_2$, $\gamma_1=\gamma_2$. 

The first theoretical predictions of optical modes with infinite radiation lifetime in periodic photonic structures were done in late 1900s -- early 2000s~\cite{bonnet1994guided, paddon2000two,pacradouni2000photonic,ochiai2001dispersion,yablonskii2002optical,tikhodeev2002quasiguided,fan2002analysis,shipman2005resonant}. At that time relation between the predicted nonradiative states and BICs with their unique properties as well as their formation mechanism was not known. The first perceived analysis of optical BICs was presented in 2008 for two arrays of identical dielectric waveguides~\cite{Marinica-2008-PRL} and for waveguide arrays with defects~\cite{bulgakov2008bound}. While disappearance of radiation was reported in very early experimental works~\cite{fox1975symposium,yablonskii2001polariton,christ2003waveguide}, the systematic experimental analysis of BICs was performed only after 2011, first in a system of coupled waveguides~\cite{Plotnik-2011-PRL}, and later for dielectric photonic crystal slabs~\cite{lee2012observation,hsu2013observation}. Further progress on theoretical and experimental studies of BICs in photonic structures up to early 2016 was summarized in the review paper~\cite{Hsu-2016-NRM}. 

Here, we briefly overview the advances in the field of photonic BICs of the last few years. As was mentioned in the introduction, nonradiating BICs exist only as a mathematical idealisation for lossless structures of infinite size or made of materials with extreme (zero or infinite) dielectric function. In practice, they transform into leaky modes, or quasi-BICs, which $Q$ factor is limited by imperfections and losses of different kind. The first methodical analysis of losses impact on BIC properties was done in Ref.~\cite{Sadrieva-2017-ACSP} where the effects of surface roughness and energy leakage into the substrate were discussed. Later, the destructive effects of the sample finiteness~\cite{bulgakov2016transfer,bulgakov2017lightenh,bulgakov2017trapping,bulgakov2018nearly,belyakov2018experimental} and disorder losses~\cite{ni2017analytical,bulgakov2017trapping,chen2018nearly,jin2018topologically} were demonstrated.

In 2014 it was shown~\cite{Zhen-2014-PRL}, that BICs supported by dielectric photonic crystal slabs demonstrate topological stability against perturbations, preserving the structural symmetry. Moreover, BICs were shown to be vortex centers in the polarization directions of far-field radiation, which carry conserved and quantized topological charges. Recently, similar analysis was done for periodic chains of dielectric rods~\cite{bulgakov2017bound} and spheres~\cite{Bulgakov-2017-PRL}. Later, polarization vortices of BICs were observed experimentally~\cite{bahari2017integrated,Doeleman-2018-NP,zhang2018observation}. Based on these findings, topologically enabled ultra-high-Q resonances were proposed~\cite{jin2018topologically} arising when multiple BIC merge in the momentum space, with the measured $Q$ factor of $5\times 10^5$. Also, topological effects were shown to allow for complete polarization conversion~\cite{guo2017topologically} at isolated wave vectors corresponding to BICs.

Recently, symmetry-protected BIC were linked to many interesting phenomena demonstrated for metasurfaces composed of arrays of meta-atoms with broken in-plane inversion symmetry, which all show the excitation of high-Q resonances for the normal incidence of light~\cite{koshelev2018asymmetric}. In particular, asymmetry of the unit cell was shown to distort a symmetry-protected BIC transforming it into a quasi-BIC with the $Q$ factor depending on the asymmetry magnitude. The proposed approach paves the way for smart engineering of resonances in metasurfaces for many applications relying on strong nonlinear response, compact platforms for sensing, active media, and electromagnetically induced transparency. We analyse the applications of broken-symmetry metasurfaces in detail below.

New twist of activity on photonic BICs is related to the recent discovery of a way to realize quasi-BICs in a single subwavelength high-index dielectric resonator. Earlier, a subwavelength core-shell particle made of material with either infinite or zero permittivity was proposed to realize a BIC at subwavelength scales~\cite{silveirinha2014trapping,monticone2014embedded}; however, for optical frequencies, such materials are not yet common, which limits the scope of possible applications. Also, we note that the earlier studies of high-Q 2D microcavities in the regime of avoided resonance crossing~\cite{wiersig2006formation} can also be linked to the BIC concept.
In the very recent pioneering work~\cite{Rybin-2017-PRL}, Rybin and co-authors proposed a novel class of high-Q nanoresonators realised by tuning the structure parameters into the so-called supercavity regime~\cite{Rybin-2017-N,taghizadeh2017quasi}, which is based on the destructive interference of several leaky modes according to the Friedrich-Wintgen mechanism. Later, strong mode coupling regime at the quasi-BIC conditions was observed experimentally in microwaves~\cite{bogdanov2019bound}. Importance of quasi-BICs for meta-optics and nanophotonics, especially, for boost of nonlinear device performance was summarized in the recent paper~\cite{koshelev2018meta}, while we overview a broader scope of applications further.

Along with the studies of field enhancement, light trapping, and topological protection governed by BICs in photonic structures, some extended research was done in this field recently, including analysis of BICs enabled due to structural anisotropy~\cite{gomis2017anisotropy,mukherjee2018topological}, self-modulation and bistability effects~\cite{pichugin2016self,yuan2017strong,KrasikovP-2018-PRB}, light-matter interaction of BICs and excitons~\cite{koshelev2018strong}, tunability of BICs~\cite{Ni-2016-PR, Timofeev-2018-PRB}, surface BICs~\cite{hsu2013bloch, tasolamprou2017near}, dynamic BICs~\cite{fan2019dynamic}, BICs in zero-index materials~\cite{silveirinha2014trapping,monticone2014embedded,lannebere2015optical,liberal2016nonradiating,li2017optical,minkov2018zero} allowing for enhancement of the photonic spin hall effect~\cite{jiang2018enhancement}, or BICs with high angular momentum in dielectric chains of rods~\cite{bulgakov2017trapping}.

\subsection*{Differences and similarities}

In this section we analyse the relationship between anapoles and BICs. For any photonic structure BICs represent eigenmodes of the system, which is not the necessary condition for observation of anapoles~\cite{Totero-2017-NC}. The problem with interpretation of anapoles arises because of their unambiguous definition in optics. 

One way to treat anapoles is to define them as special localized distributions of electromagnetic field which are manifested as the local minima of the scattered power spectrum. For lossless 2D and 3D structures, such as infinite dielectric rods or isolated nanoparticles, the scattered power $P_{\rm sca}$ can be rigorously written as a sum of Fano profiles describing the resonant response in the vicinity of eigenmodes (see Ref.~\cite{Totero-2017-N} and Supplemental materials for Ref.~\cite{bogdanov2019bound})
\begin{align}
&P_{\rm sca}(\omega) = \sum_n \alpha_n P_{n}(\omega) + P_{\rm bg},\\
&P_{n}(\omega) = \frac{1}{1+q_n^2}\frac{\left[q_n+\frac{2(\omega-\omega_n)}{\gamma_n}\right]^2}{1+\left[\frac{2(\omega-\omega_n)}{\gamma_n}\right]^2},
\label{eq:Fano}
\end{align}
where $\alpha_n$ and $P_{\rm bg}$ is the non-resonant envelope and background contribution, respectively, $P_{n}$ is the scattering Fano profile of $n$-th eigenmode, the summation goes over the eigenmodes with resonant frequencies $\omega_n$ and inverse radiation lifetimes $\gamma_n$ producing peaks in the scattering spectrum with the Fano asymmetry parameter $q_n$.
At the same time, for lossless periodic photonic structures we can define the total scattered power as
\begin{equation}
P_{\rm sca}/P_0 =  |r|^2+|1-t|^2 = 2(1-|t|\cos{\phi_t}),
\label{eq:scsPer}
\end{equation}
where $r$ and $t$ are the complex amplitudes of reflected and transmitted fields, $\phi_t$ is the phase of $t$, and $P_0$ is the incident power. Here, $t$ and, consequently, $P_{\rm sca}$ can be also rigorously decomposed into a series of Fano profiles (see Supplemental materials for Ref.~\cite{koshelev2018asymmetric}).

The analysis of Eqs.~(\ref{eq:Fano}) and ~(\ref{eq:scsPer}) shows that the local minimum of $P_{\rm sca}$ can be achieved for both resonant and non-resonant conditions. In the first case, excitation of peculiar resonances with a pump of a special geometry can lead to dips in the scattering spectrum characterized by the asymmetry parameter $q_n =0$ which are treated as anapole states, representing the true eigenmodes of the structure. Such states were recently predicted for isolated high-index disks~\cite{Rybin-2017-PRL,bogdanov2019bound}. In the non-resonant scenario the interplay between several resonances with different frequencies can also lead to the scattering minimum, describing the formation of an anapole being not an inherent state of the resonant structure but a dynamic excitation which depends strongly on external parameters such as properties of the incident beam or the excitation geometry. Importantly, quasi-BICs supported by finite optical structures always produce symmetric Lorentzian peaks in the scattering spectrum of plane waves, which is characterized by pronounced maxima~\cite{bogdanov2019bound}, differing substantially from spectral dips, originating from anapole behaviour.

Moreover, Eq.~(\ref{eq:scsPer}) shows that the anapole conditions in periodic structures imply the light transmission through a resonant system without perturbation of both amplitude and phase $|t|=1$, $\phi_t=0$, which could be referred to a lattice invisibility effect~\cite{terekhov2019Multipole,sayanskiy2018controlling}, when electromagnetic energy is accumulated in scatterers but the total electromagnetic fields outside the scatterers mimic the incident fields without distortion. On the contrary, since BICs possess infinite $Q$ factors, they cannot couple to radiation channels and accumulate energy when excited from the far field.

Another way to recognize anapoles in optical structures is to link them with destructive interference of toroidal and dipolar modes, or more specific, define them as dynamic current distributions characterized by the compensation or absence of the leading multipole order. Then, for specific cases, when leading multipole term is of dipolar nature, its suppression can be explained as complete compensation of a dipole and toroidal contributions. In comparison to anapoles, for BICs the radiation is completely prohibited, which can be achieved only by destructive interference of all multipoles at the same time.


\section*{Applications}

Controlling localized electromagnetic fields of high intensity is a desired goal in nanotechnology. Plasmonic materials opened up an impressive number of applications that take advantage from localized energy hot-spots employed for bioimaging, ultrafast dynamics, energy harvesting, catalysis and enhanced fs-chemistry reactions \cite{Maier:2007:10.1007/0-387-37825-1,mosk2013,fra_2017,zay}. These results, however, come with a limitation originated from the high losses of metallic structures. This problem is currently at the center of extensive research that aims at creating new materials able to overcoming this issue~\cite{Hess-2012-NM}.

As we discuss below, nonradiating states of light lead to the generation of strongly localized electromagnetic fields, similar to their plasmonic counterparts, but in all-dielectric nanomaterials with negligible or very limited losses, and as such they hold the promise of extending the results of plasmonics into new applications where minimal losses are required. In addition, due to the multipolar nature and interferences associated with such states, all-dielectric platform brings many novel effects not even possible in plasmonics mainly driven by the physics of electric hot-spots and electric dipole modes. 

%
\subsection*{Field enhancement}

\begin{figure}[t]
  \centering
  \includegraphics[width=\columnwidth]{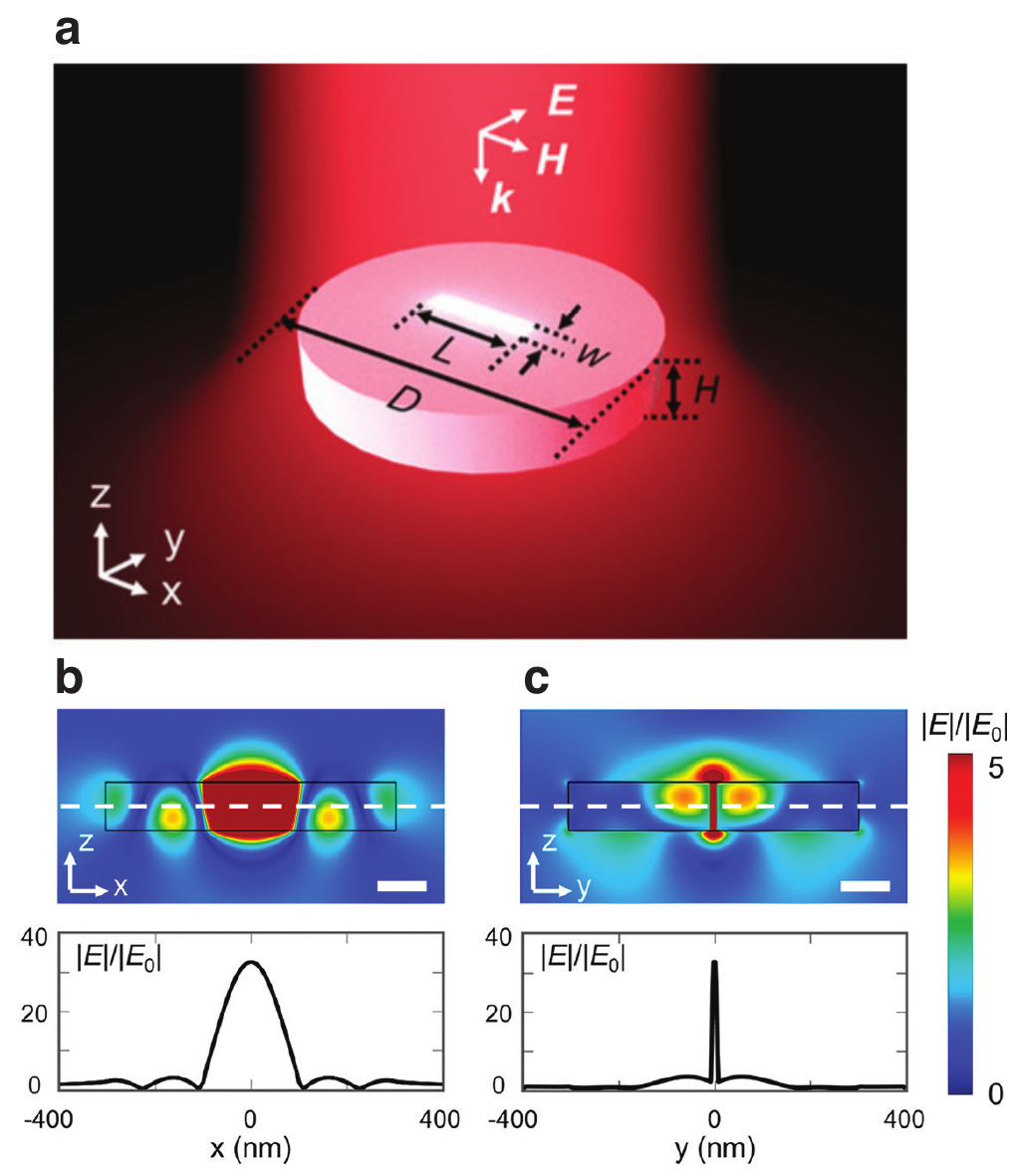}
  \caption{
      \label{figD}
 				\textbf{Anapole in a slotted nanodisk for enhanced light confinement.}
        (a) Structure composed of a slotted Si nanodisk. This system is able to sustain the anapole state, while benefiting from advantageous boundary conditions to generate a highly intense and concentrated hot-spot at the center. (b-c) Near-field enhancement of the electric field in the nanodisk structure. Adapted with permission from \cite{Yang-2018-ACSP}.
	}
\end{figure}


The strong field enhancement up to several times localized in the center of the resonator was obtained due to excitation of the anapole state in isolated nanoparticles~\cite{Miroshnichenko-2015-NC}. This enhancement is governed by the interference between the electric and toroidal dipoles, which cancel each other far away from the structure but mutually enhance the fields inside the resonator. 

The higher field enhancements, up to two orders of magnitude of the electric field, were reported for individual slotted nanodisks~\cite{Yang-2018-ACSP}. The enhancement was improved by taking advantage of high refractive index of the material supporting the anapole state, and, in particular, by engineering a suitably designed air hole in the nanodisk. Field enhancement in high-index dielectric metasurfaces with gapped meta-atoms at the anapole conditions was analysed in Ref.~\cite{Liu-2017-OE}, where the values of enhancements of up to several thousands were reported. All-dielectric nanostructure configurations with air gaps are of particular interest in spectroscopy and sensing applications, as they generate energy hot-spots localized outside of the dielectric structure, and are expected in future to stimulate applications of anapole states for those important applications.

\begin{figure*}[tbp]
  \centering
  \includegraphics[width=16cm]{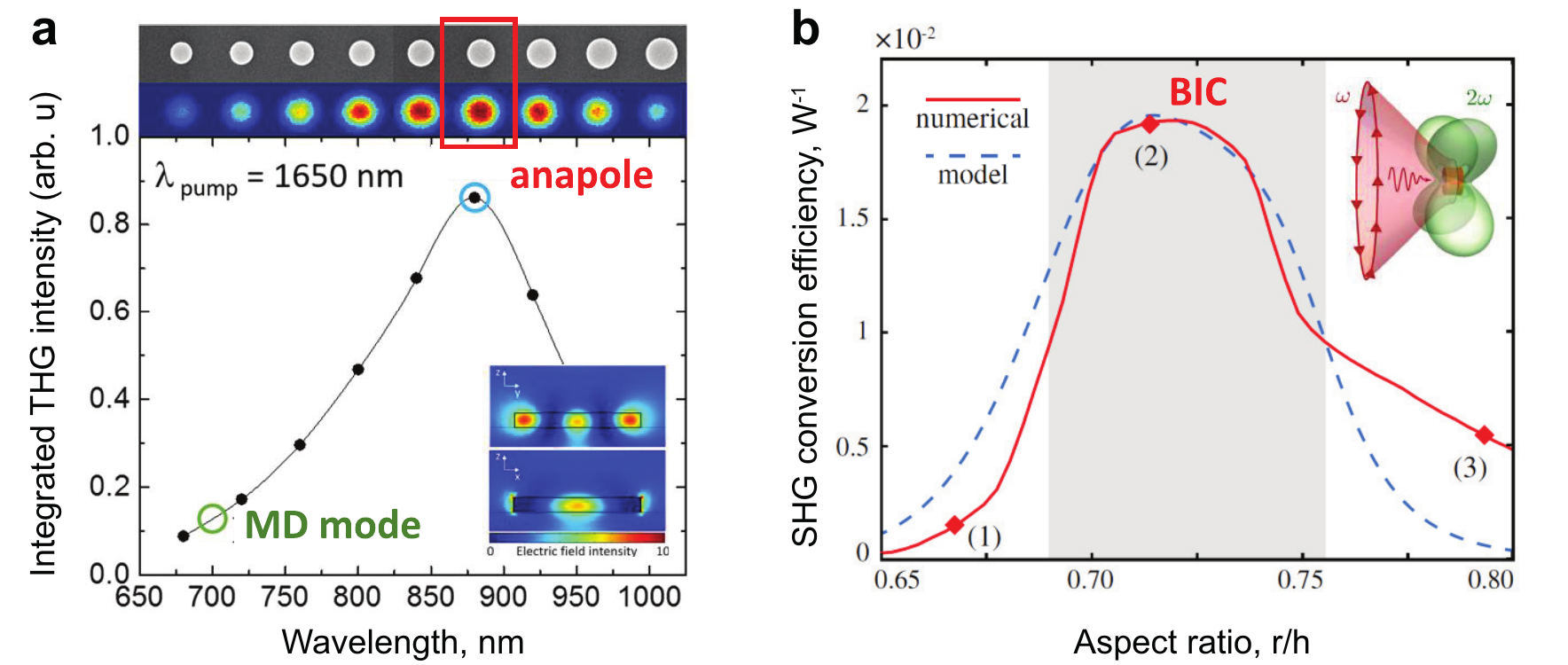}
  \caption{
    \label{fig:SHG}
        \textbf{Enhanced harmonic generation.}  (a)  Observation of the enhanced nonlinear THG in Ge nanodisks near the anapole conditions. Insert shows the numerically simulated field structure of the anapole. Adapted with permission from~\cite{Grinblat-2016-NL}. (b) Predicted giant SHG near the BIC conditions in AlGaAs a nanorod. Insert shows the multipolar field structure. Adapted with permission from~\cite{Luca_PRL_2018}.}
\end{figure*}


Another important aspect closely related to field enhancement is the quality factor ($Q$ factor), which governs the performance of any resonating structure~\cite{ref999550742402121}. Importantly, since $Q$ factor is the resonant property of any cavity, it can be correctly defined only for anapoles representing eigenstates of the resonator. The idea of using an anapole state to increase the $Q$ factor of a resonator structure was first exploited in metamaterials in the GHz frequency range, with an array of double split-ring metallic meta-atoms~\cite{Basharin-2017-PRB}. Thanks to the generation of a nonradiating anapole state, this approach reported extremely high $Q$ factor (about $10^6$) and near-field enhancements (about $10^4$). The range of operation of this metasurface, however, does not cover the optical range completely.

The first structure operating in the optical range and relying on dielectrics was reported with split nanodisks arrays~\cite{Liu-2017-OE}, which represents a  generalization of the structure introduced in~\cite{Basharin-2017-PRB}. Despite low $Q$ factor of each nanodisk composing the array, the collective excitation of anapoles through near field couplings reported a significant enhancement of the $Q$ factor of the overall structure, which reached values of $10^6$ with two orders of magnitude field enhancement in the near-field inside each nanodisk.

The field enhancement and increase of the $Q$ factor for BICs was also analysed in photonics. In the papers~\cite{mocella2015giant, yoon2015critical} giant field enhancement was reported for periodic gratings and photonic crystal slabs. For individual  resonators, the dramatic enhancement of the $Q$ factor was proposed recently by Rybin and co-workers~\cite{Rybin-2017-PRL}, who reported theoretically quasi-BICs in isolated subwavelength silicon nanodisks with $Q$ factors of order of $200$ operating in the optical range. The demonstrated high-Q resonances were shown to originate in accord with the Friedrich-Wintgen mechanism which predicts the strong mode coupling and almost complete destructive interference of several leaky modes in the far field provided the resonator parameters are tuned appropriately. More recently, the physics of strong mode coupling, origins of radiation suppression and relation to the peculiarities of Fano resonances were analysed for such quasi-BICs in subwavelength high-index structures~\cite{bogdanov2019bound}, supported by the experimental verification in microwaves of the characteristic avoided resonance crossing behaviour at the BIC condition. Also, the formation of high-Q states was explained in terms of distinct changes in the multipole decomposition in the vicinity of quasi-BICs accompanied by drastic change of the far-field radiation pattern~\cite{bogdanov2019bound,chen2018subwavelength}. Later, the enhancement of the Purcell factor at the quasi-BIC regime was reported~\cite{wei2018dissipative}.

\subsection*{Nonlinear harmonic generation}

High $Q$ factors and strong field enhancement sustained by nonradiating optical states are of natural interest for nonlinear applications for enhancing the efficiency of nonlinear light-matter interaction. As remarked in Ref.~\cite{Ospanova-2018-LPR}, dielectric materials such as silicon have great interest in those applications due to their strong thermal stability and high-intensity threshold for optical damage.

A widely exploited nonlinear effect is the higher harmonics generation and multi-frequency wave mixing, which include nonlinear wave generation at combined frequencies (sum, multiple, or difference) of the frequencies of the incoming waves.

The first approach exploiting the anapole state to enhance nonlinear harmonic generation was proposed in Germanium nanodisks (of $\SI{100}{nm}$ of thickness and radii ranging from 250 to $\SI{500}{nm}$) by Grinblat {\em et al.} in Ref.~\cite{Grinblat-2016-NL} (see Fig.~\ref{fig:SHG}a). When excited in the vicinity the anapole resonance, taking advantage of a high field confinement, this structure performed a third-harmonic conversion from incident infrared light ($\SI{1650}{nm}$) into green light ($\SI{550}{nm}$) with an efficiency that is orders of magnitude higher with respect to a flat structure.

In Ref.~\cite{Grinblat-2017-ACSN}, the authors showed that higher-order anapoles supported by larger nanodisks resulted in narrower scattering cross-section and  higher localized field confinement. This generated a conversion efficiency one order of magnitude higher than that observed in~\cite{Grinblat-2016-NL} for third-harmonic generation (THG) at $\SI{500}{nm}$.\\
The slit nanodisk array configuration introduced by Liu and co-workers in~\cite{Liu-2017-OE} and described above also offered enhanced THG conversion efficiency comparable to that of Ref.~\cite{Grinblat-2017-ACSN}, without resorting to high-order anapoles but by exploiting an optimized fundamental anapole.

Another path to boost even more the \ac{THG} efficiency was reported by Shibanuma and co-workers in~\cite{Shibanuma-2017-NL}. This work leveraged on the coupling of anapoles sustained in dielectric nanodisks with metallic nanostructures. A conversion efficiency of 0.007\% (seven times higher than in ~\cite{Grinblat-2017-ACSN} and ~\cite{Liu-2017-OE}) at a third-harmonic wavelength of $\SI{440}{nm}$ was obtained by adding a metallic annulus around a silicon nanodisk. The highest THG conversion efficiency of 0.01\% was expectantly achieved by placing a silicon nanodisk on top of a gold reflector, benefiting from free-charge oscillations at the interface~\cite{Xu-2018-LSA}.

\begin{figure*}
  \centering
  \includegraphics[width=17cm]{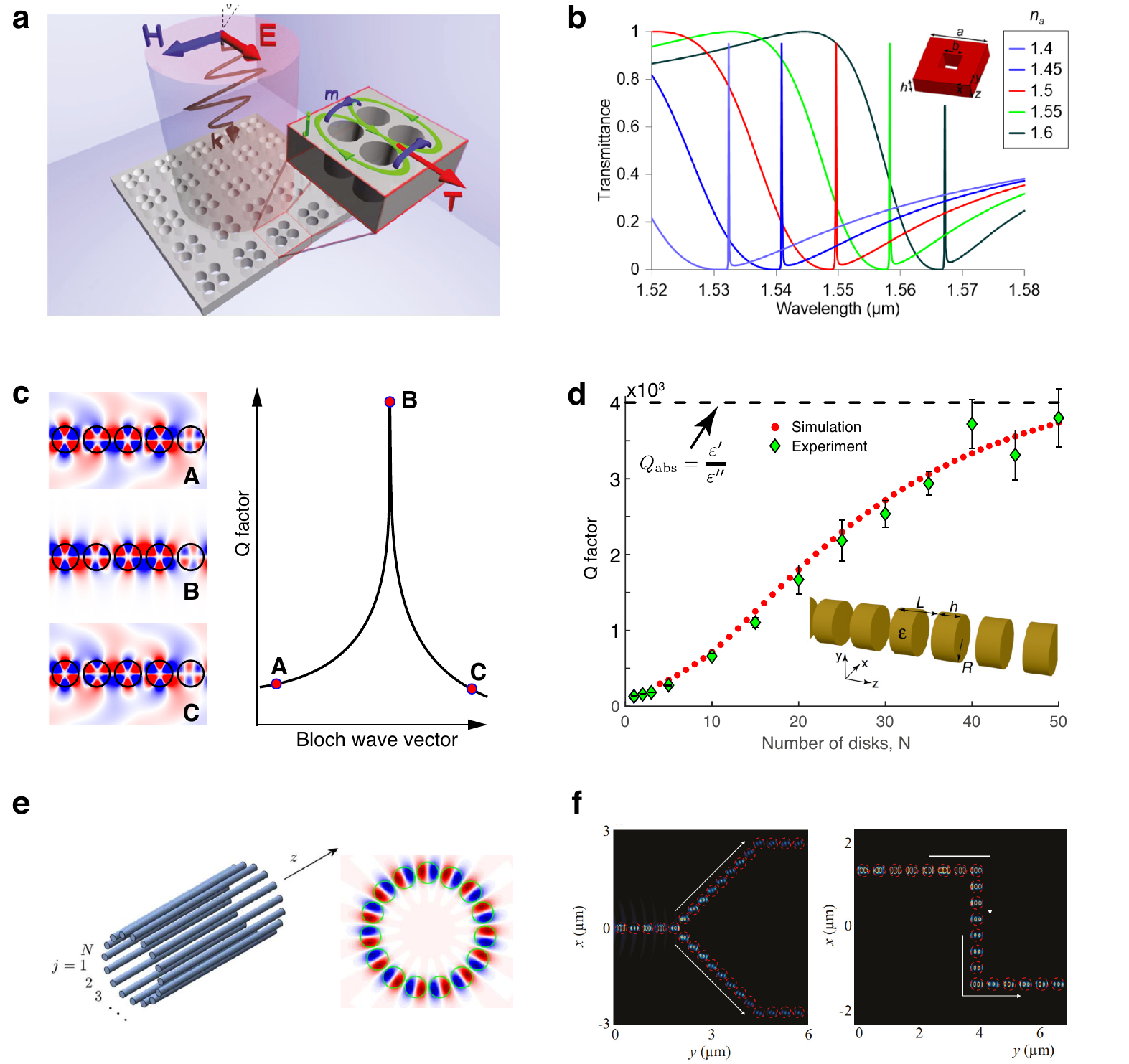}
  \caption{
  \label{fig_chains_and_MS}
  \textbf{Anapole and BIC metasurfaces and metamaterials.}
(a) Illustration of the metamaterial, composed of planar \SI{100}{nm} thick Si layer with cylindrical perforations, supporting anapole modes in the near-infrared range. 
(b) Transmittance spectra of high-refractive-index hollow cuboid metasurface supporting anapole mode depending in refractive index of the analyte superstratum. The inset shows the design of the unit cell.  
(c) The characteristic dependence of the radiating $Q$ factor of the leaky mode  with zero orbital momentum in a chain of dielectric spheres on the Bloch wavenumber.  Panels {\bf A} and {\bf C}  show distribution of azimuthal component of the electric field of the leaky modes. Panel {\bf B} shows distribution of azimuthal component of the electric field of the accidental BIC.
(d) Dependence of the radiating $Q$ factor of the symmetry protected BIC in chain of the ceramics disks on number of the disks. 
(e)  Novel type of the waveguide consisting of a circular array of parallel dielectric wires providing optical signals above the light line of the surrounding space without radiation losses. 
(f)  Dielectric disks arranged to Y-shaped splitter and S-shaped bend waveguide providing efficient transfer of anapole excitation.  
Adapted with permission from~\cite{Ospanova-2018-LPR,algorri2019anapole,bulgakov2017light,belyakov2018experimental,Mazzone-2017-AS,bulgakov2018fibers}.
	}
\end{figure*}
Other than THG, enhanced second-harmonic generation (SHG) was also recently reported by using nonradiating anapoles by Timofeeva and co-workers in Ref.~\cite{Timofeeva-2018-NL}. This work proposed an interesting fabrication process relying on ion-beam billing to slice nanowires, resulting in nanodisks in a vertical free standing configuration. This approach can have significant interest for a design of arrays and lattices of nanodisks. 

Recently, a smart route to enhance the efficiency of nonlinear harmonic generation in metasurfaces was suggested relying on quasi-BICs in periodic arrangements of asymmetric meta-atoms, which are analysed in the next section. We also mention here the recent papers~\cite{wang2017improved,wang2018large} devoted to investigation of the enhancement of SHG and THG by BICs in periodic structures coupled to a graphene sheet and a transition metal dichalchogenide layer. 

Complex wave mixing effects were also considered by using non radiating states in~\cite{Grinblat-2017-ACSP}, which reported \ac{FWM} in a similar structure of~\cite{Grinblat-2016-NL}, and composed of a germanium nanodisk ($\SI{200}{nm}$ of thickness and $\SI{625}{nm}$ of radius) in which two incident waves at different frequencies ($\omega_1$ and $\omega_2$) give rise to four waves ($3\omega_1$, $2\omega_1+\omega_2$, $\omega_1+2\omega_2$, $3\omega_2$) with enhanced efficiency. Moreover, the frequency comb generation in two coupled resonators supporting a BIC was shown to provide a way for all-optical generation of waves with ultralow frequency~\cite{pichugin2015frequency}.

The study of nonlinear effects with isolated nanoparticles has been initiated only very recently, with the experimental demonstrations yet to come. Carletti {\em et al.}~\cite{Luca_PRL_2018} studied SHG from isolated subwavelength AlGaAs nanoparticles and predicted that nonlinear effects at the nanoscale can be boosted
dramatically provided the resonator parameters are tuned to the BIC regime. They predicted a record-high conversion efficiency for nanoscale resonators that exceeds by two orders of magnitude the conversion efficiency observed at the magnetic dipole Mie resonance (see Fig.~\ref{fig:SHG}b), thus opening the way for highly efficient nonlinear metasurfaces and metadevices.  A brief summary of recent advances in this field was reported in Ref.~\cite{koshelev2018meta}.

In addition, we also mention that high Q factors at the
nanoscale can boost other nonlinear processes, such as the spontaneous parametric downconversion. Recently, the quasi-BIC supported by the AlGaAs nanodisk was shown to produce a strong enhancement of generation of entangled photon pairs associated with electric and magnetic dipole modes~\cite{poddubny2018nonlinear}.

\subsection*{Metamaterials and metasurfaces}


To bring the unique properties of anapoles to the macroscopic scale, the resonators supporting anapole states could be arranged into periodic arrays. The toroidal dipole moment in metasurfaces was analyzed theoretically and experimentally at different frequency ranges, from microwaves to the visible spectrum  (see for example~\cite{kaelberer2010toroidal,gupta2016sharp,dong2012optical,liu2017high,huang2012design}). The majority of considered designs contains metallic parts, which do not give essential contribution to losses in microwave and THz regions because of extremely small skin depth at such frequencies. However, the losses in metallic parts plays the dominant role in the visible and near infrared regions. Thus, the performance of plasmonic metamaterials with toroidal dipole moment at the visible frequencies is not so high as in the radio-frequency range~\cite{li2014excitation,dong2012optical}. Dielectric metamaterials are free from this disadvantage. It was shown theoretically and experimentally in Ref.~\cite{xu2018experimental} that the dielectric metasurface consisting of trimer clusters of high-index dielectric particles demonstrate strong toroidal response in GHz frequency range in the near- and far-field region. This design can be scaled for measurements in the visible range. Other designs of all-dielectric metasurfaces with strong toroidal response for the visible range were proposed in Refs.~\cite{sayanskiy2018all,tuz2018all}.

Metasurface can support anapole excitations if the toroidal moment of the unit cell is compensated by the dipole moment. Such type of metasurfaces are under active study today. The anapole mode in metasurfaces being a high-Q resonance manifests itself in transmission spectrum as an ultra narrow dip, which was confirmed experimentally in the GHz frequency range~\cite{Basharin-2017-PRB}. The metasurface consists of metamolecules formed by the two symmetrical split rings. The incident wave linearly polarized along the central wire induced the in-plane toroidal moment which is compensated by the electrical dipole moment induced in the gaps of the metamolecule. The experimentally demonstrated $Q$ factor is of $8.5\times10^4$ at 10 GHz. The design of all-dielectric metasurface supporting the high-Q anapole mode in the near-infrared range was proposed in Ref.~\cite{Liu-2017-OE}. The metasurface consisting of silicon split nanodisk arrays demonstrates (in the numerical experiment) $Q$ factor approaching $10^6$ and enhancement of the electric field in the gap of about 800. The similar metasurface but consisting of split Si nanocuboids allows to achieve (in the numerical experiment) the $Q$ factor of $4\times10^6$ and electric field enhancement of 1450 in the near-infrared range~\cite{sun2018strong}. 

\begin{figure*}[t]
  \centering
  \includegraphics[width=15cm]{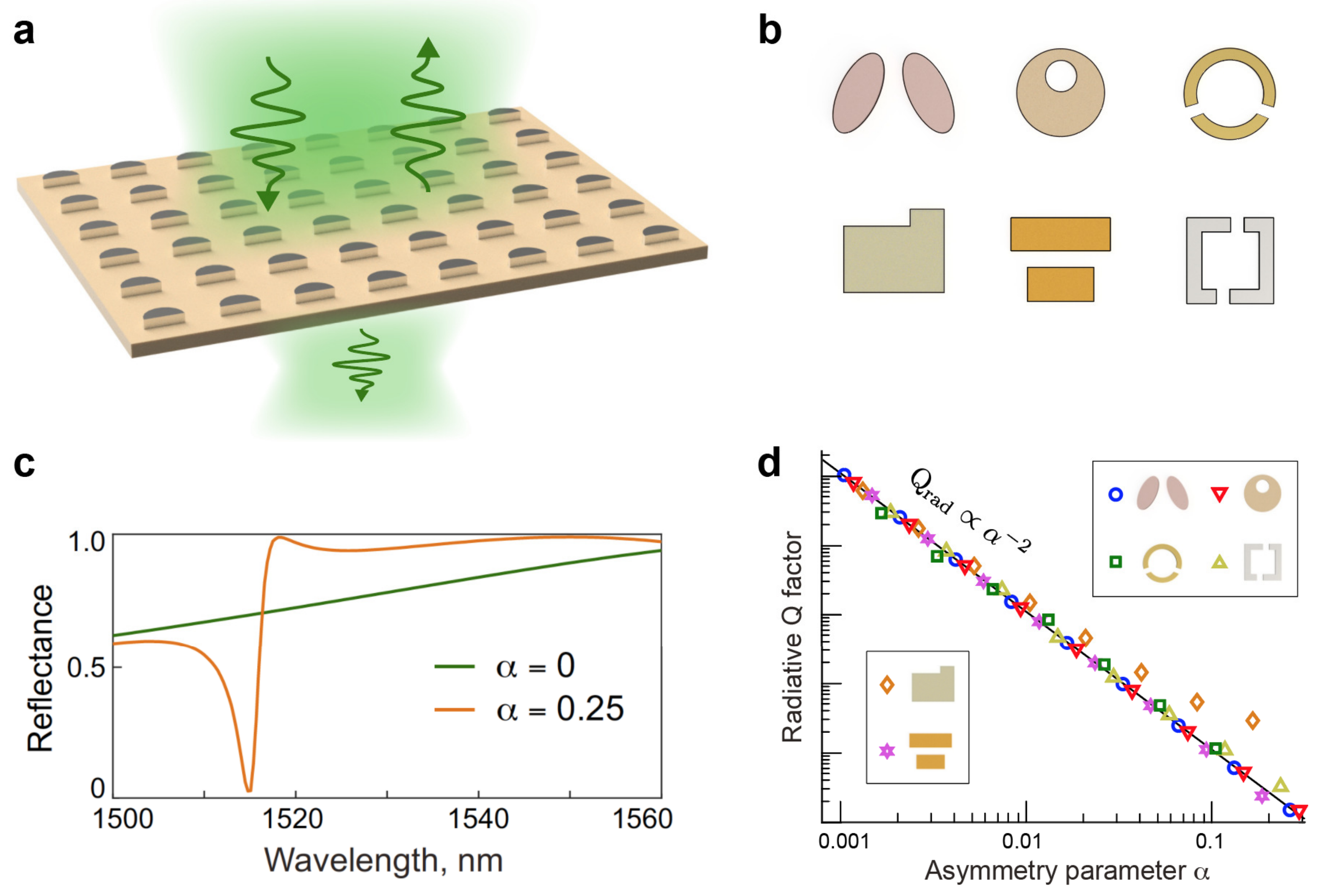}
  \caption{
      \label{fig:asym}
        \textbf{BICs in metasurfaces with a broken symmetry.}   (a) Schematic of the scattering of light by a metasurface. (b) Designs of the unit cells of asymmetric metasurfaces with a broken in-plane inversion symmetry of constituting meta-atoms supporting sharp resonances, as considered in Refs.~\cite{tittl2018imaging,tuz2018high,fedotov2007sharp,vabishchevich2018enhanced,forouzmand2017all,lim2018universal,singh2011observing}. (c) Example of evolution of the reflectance spectrum of the metasurface from the symmetric geometry ($\alpha$ = 0) to a broken-symmetry geometry ($\alpha$ = 0.25). Here, $\alpha$ is the asymmetry parameter.  (d) Dependence of the Q factor on the asymmetry parameter $\alpha$ for all designs of symmetry-broken meta-atoms shown in (b)  (log-log scale). Adapted with permission from Refs.~\cite{koshelev2018asymmetric,koshelev2018meta}.	}
\end{figure*}


The idea to exploit the high-Q anapole mode in all-dielectric metasurfaces for sensing was developed in Ref.~\cite{algorri2019anapole}. The Authors analyzed numerically the shift of the anapole mode in high-refractive-index hollow cuboids arranged in a planar metasurface due to the analyte superstratum (Fig.~\ref{fig_chains_and_MS}b). The metasurface with optimized design exhibits the $Q$ factor of $1.7\times10^6$ and sensitivity of up to 180 nm/RIU. However, it is still lower than sensitivity of the commercially available surface plasmon resonance sensors based on Kretschmann configuration, which is about 10$^3$-10$^4$~nm/RIU~\cite{homola1999surface}.     

The design of a metasurface supporting anapole modes in the visible range was proposed recently by Ospanova and co-workers~\cite{Ospanova-2018-LPR}, relying on a metasurface composed of a perforated \SI{100}{nm} thick Si layer and represented in Fig.~\ref{fig_chains_and_MS}a. The unit cell cell of this structure is composed of 4 holes of \SI{45}{nm} radius and distant of $\SI{55}{nm}$ arranged in an array with a periodicity of $\SI{200}{nm}$. This metamaterial structure features a full transmission peak for a given frequency (\SI{566.5}{THz} in this design) that is controlled through the depth of the holes. The metamaterial is easily fabricated in one step through the \ac{FIB}.

The dynamic anapole states could be transferred through the chain of the coupled resonators. The chain of dielectric disks guiding the anapole excitation was proposed by Mazzone and co-workers in~\cite{Mazzone-2017-AS}. It was shown that such waveguides are extremely robust against physical bending and splitting, and that the anapole state is efficiently transmitted through the S-shaped bend waveguide and Y-shaped splitter (Fig.~\ref{fig_chains_and_MS}f). Thanks to the near-field confinement, this type of approach also allowed for the integration of all dielectric optical circuitry with a high density.

Gathering of single resonators into periodic arrays results in appearance of collective interference phenomena not inherit to single resonators. For instance, all multipole moments (except anapoles) in single resonators radiate to far-field. In a periodic array due the interference between all meta molecules, the radiation occurs only to a finite number of allowed directions -- the open diffraction channels. Therefore, in periodic structure any multipole can be non-radiating if the directions of open diffraction channels coincide with the nodes of far-field radiation pattern of the multipole. This simple physics explains appearance of BICs in periodic photonic structures. Therefore, BICs can transfer energy without radiation losses in periodic structures (gratings, chains, metasurfaces) even if there are open diffraction channels~\cite{bulgakov2018propagating, Hu-2017-JOSA,yuan2017propagating,bulgakov2017propagating,bulgakov2018fibers}. However, it is possible only at certain frequencies. 

Figure~\ref{fig_chains_and_MS}c shows the dependence of the radiating $Q$ factor of the leaky mode in the 1D array of dielectric spheres on the Bloch wavenumber analyzed in Ref.~\cite{bulgakov2017light}. One can see that radiation losses disappear only at the certain value of the Bloch wavenumber, when the leaky mode turns into the accidental BIC with non-zero group velocity. The insets in Fig.~\ref{fig_chains_and_MS}c show the field distribution of the leaky modes and BIC. Such type of chains were analyzed theoretically in details in Refs.~\cite{Bulgakov-2017-PRA,bulgakov2017bound,polishchuk2017guided,bulgakov2015light,bulgakov2017light,Bulgakov:18,Bulgakov-2017-PRL,han2018trapped,bulgakov2017trapping}. In particular, A.~Sadreev and co-workers proposed a novel type of the waveguide consisting of a circular array of parallel dielectric wires providing optical signals above the light line of the surrounding space without raidiation losses (Fig.~\ref{fig_chains_and_MS}e)~\cite{bulgakov2018fibers}. 

\begin{figure*}[t]
  \centering
  \includegraphics[width=13cm]{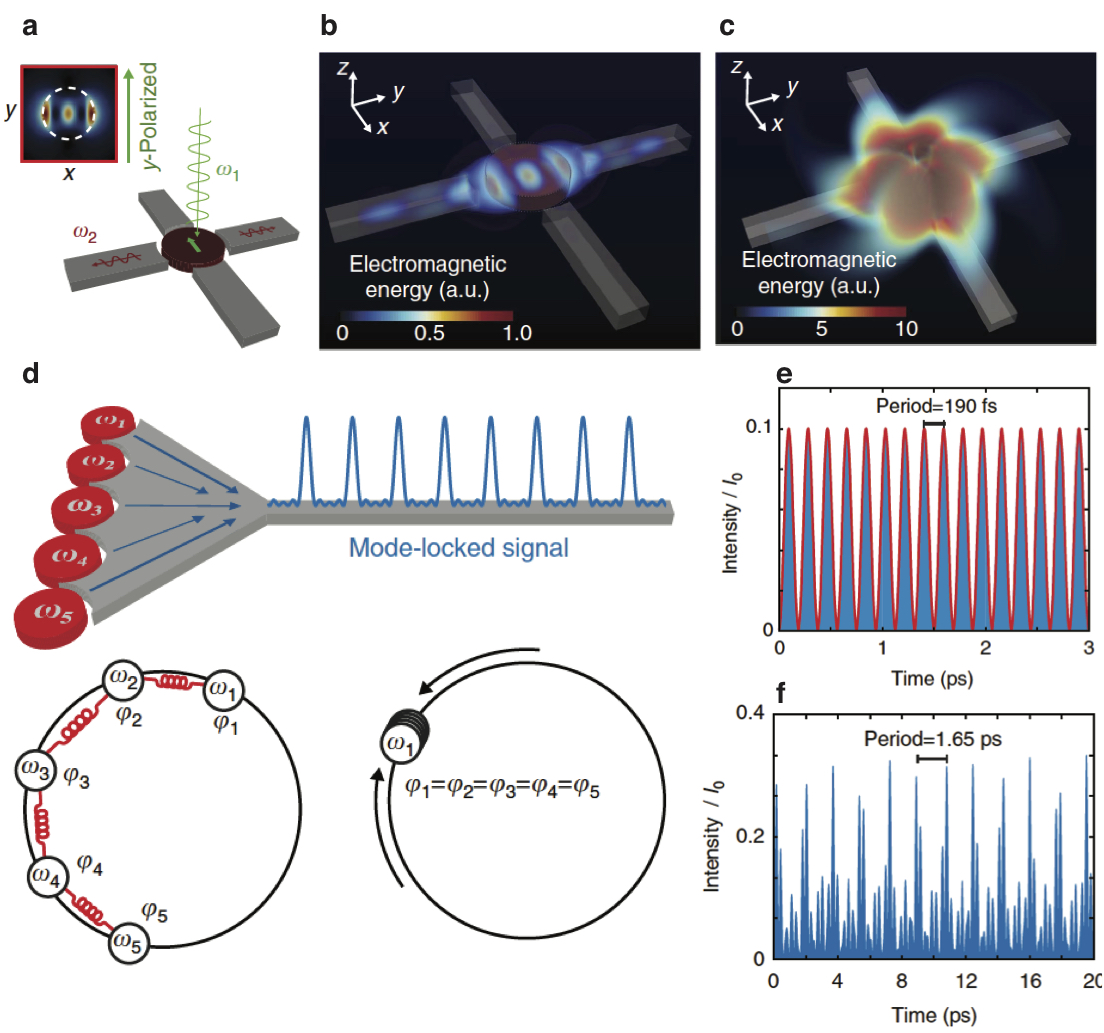}
  \caption{
      \label{fig6}
        \textbf{Concept of an anapole nanolaser.}
        (a-c) Anapole router. (a) An InGaAs nanodisk is placed in the middle of an optical circuit composed of four silicon nanowire waveguides disposed in a cross configuration around the nanodisk. (b) By pumping linearly polarized light onto the nanodisk, the anapole mode can be selectively
        excited within the nanodisk: only the anapole whose dipolar field is perpendicular to the pumped light polarization is excited. (c) Comparison by pumping a scattering mode of the InGaAs structure: no coupling occurs.
        (d-f) Ultrafast pulsed source on a photonic chip. (d) An anapole chain of weakly coupled nanodisks generates a mode-locked signal propagating in the nanowire channel as an ultrafast pulse train. (e-f) Pulse trains of different periods generated by acting on the chain geometry.
        Adapted with permission from~\cite{Totero-2017-NC}.
	}
\end{figure*}

The first experimental observation of BIC in chain of dielectric resonators was done in Ref.~\cite{belyakov2018experimental}. The Authors consider an array of ceramics disks and observed symmetry protected BIC in transmission spectra and by mapping of the near-field distribution. A true BIC exists only in the infinitely long chain when the radiation losses are completely canceled due to the destructive interference. In a finite chain, the radiation losses exist always but they can be suppressed to a level much lower than other losses in the system. In Ref.~\cite{belyakov2018experimental}, the Authors analyzed the dependence of the $Q$ factor of the symmetry protected BIC on number of the disks (see Fig.~\ref{fig_chains_and_MS}d) and show that the array consisting of several tens of the disks behaves as an infinite one.
Observation of accidental BIC and BIC with high orbital angular momentum in a chain of resonators is still an unsolved experimental challenge. 
      
Recently, the resonant response of a wide class of  asymmetric metasurfaces consisting of unit cells with broken in-plane inversion symmetry was shown to demonstrate narrow features with the Fano shape in the normal incidence reflection and transmission spectra. Such sharp resonances were treated as high-Q states and used for various applications, including the efficient imaging-based molecular barcoding~\cite{tittl2018imaging} and enhanced nonlinear response~\cite{vabishchevich2018enhanced}. In a very recent paper it was proven than all observed high-Q resonances in such seemingly different metasurfaces with broken-symmetry meta-atoms can be unified by the BIC concept~\cite{koshelev2018asymmetric}.
In particular, it was demonstrated that such resonances originate from symmetry-protected BICs transformed into quasi-BICs due to breaking of the in-plane inversion symmetry. More specifically, it was shown that the unit cell asymmetry induces an imbalance of the interference between contra-propagating leaky waves comprising the symmetry-protected BIC that leads to radiation leakage. Also, the universal inverse square law for the Q factor dependence on the asymmetry parameter was analytically and numerically proven for arbitrary metasurfaces with small and moderate degrees of asymmetry. The reported universality paves the pathway for smart desiging of efficient devices for meta-optics applications.

\begin{figure*}[t]
  \centering
  \includegraphics[width=14cm]{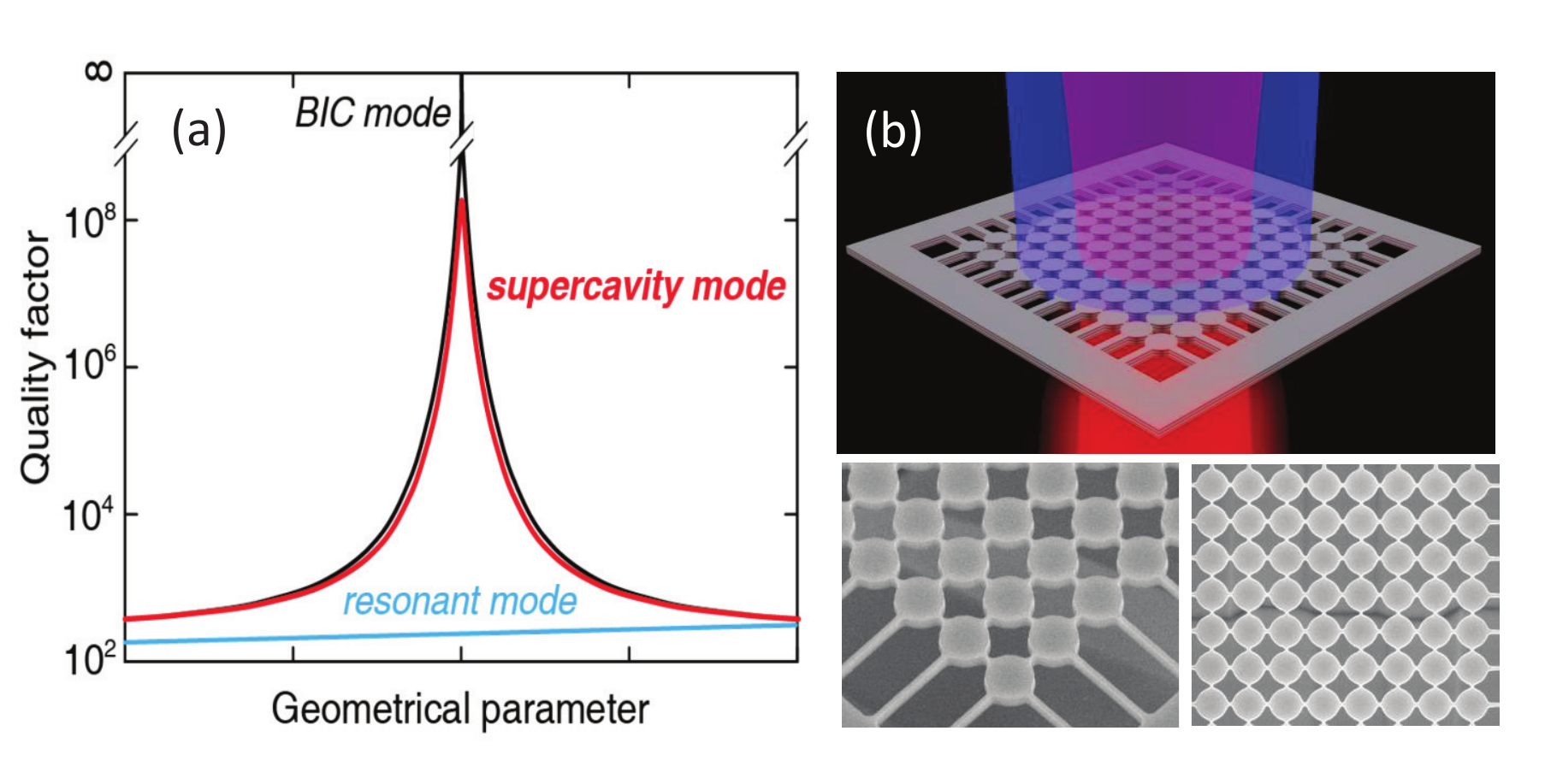}
  \caption{
      	\label{figH}
        \textbf{BIC laser.}
        (a) Schematic of the $Q$ factor dependence for lasing cavities.
        In conventional optical cavities (in blue), the $Q$ factor have a very weak dependence on geometrical parameters. BICs (in black) on another hand have a very strong dependence on geometrical parameters, and can theoretically reach an infinite $Q$ factor for an infinite structure with optimized set of parameters. Based on BICs, finite supercavities (red) can be engineered to reach finite yet extremely high values of the $Q$ factor, with an important dependence on the geometrical parameters.
        (b) Photonic cavity for lasing based on BICs proposed in~\cite{Kodigala-2017-N}.
        The cavity is composed of a lattice of $N\times N$ InGaAsP nanodisks linked by bridges and disposed into a membrane suspended in air. For an optimized set of parameters lasing action can happen within the membrane at room temperature. Adapted with permission from~\cite{Rybin-2017-N}.
	}
\end{figure*}

The findings of Ref.~\cite{koshelev2018asymmetric} are summarized in Fig.~\ref{fig:asym}. The schematic of a broken-symmetry metasurface scattering light is shown in Fig.~\ref{fig:asym}a. Figure~\ref{fig:asym}b demonstrates the examples of designs of the unit cells of asymmetric metasurfaces with a broken in-plane inversion symmetry which were reported to produce sharp resonances in reflection or transmission~\cite{tittl2018imaging,tuz2018high,fedotov2007sharp,vabishchevich2018enhanced,forouzmand2017all,lim2018universal,singh2011observing}. A typical behaviour of change of the reflectance spectrum from the metasurface with symmetric geometry ($\alpha$ = 0) to a broken-symmetry metasurface ($\alpha$ = 0.25) is shown in Fig.~\ref{fig:asym}c, where $\alpha$ is the asymmetry parameter.  Figure~\ref{fig:asym}d demonstrates the universal inverse quadratic dependence of the Q factor on the asymmetry parameter $\alpha$ for all designs of symmetry-broken meta-atoms shown in Fig.~\ref{fig:asym}b. 

We believe that asymmetric metasurfaces based on the BIC concept can describe many other phenomena studies earlier with different applications in mind, for example, the electromagnetically induced transparency~\cite{hu2018comparison}, sensing~\cite{he2013electromagnetically}, and light emission control~\cite{liu2018light}. 

\subsection*{Novel concepts for lasing}
The strong field confinement associated with anapoles and BICs stimulated a design of new types of nanolasers. The anapole concept was used by Totero and co-workers in Ref.~\cite{Totero-2017-NC}  to design a new type of nonradiating laser source, composed of a semiconductor nanodisk hosting an anapole mode and optically pumped as a conventional multilevel laser. The semiconductor nanodisk suggested in this work was made of $\mathrm{In}_{x}\mathrm{Ga}_{1-x}\mathrm{As}$ and provided stimulated emission tunable from $\SI{873}{nm}$ to $\SI{3.6}{nm}$ by varying the relative concentration of In and Ga.

The proposed idea was to use stimulated emission to amplify an anapole state, which remained confined inside the nanodisk with an emission wavelength that matches that of $\mathrm{In}_{x}\mathrm{Ga}_{1-x}\mathrm{As}$. This study~\cite{Totero-2017-NC} showed that such nonradiating source is coupled efficiently to nearby waveguide through near-field coupling, following the same approach as in the case of the anapole and BIC waveguides~\cite{Mazzone-2017-AS,Bulgakov-2017-PRL}.

\begin{figure*}[t]
  \centering
  \includegraphics[width=13cm]{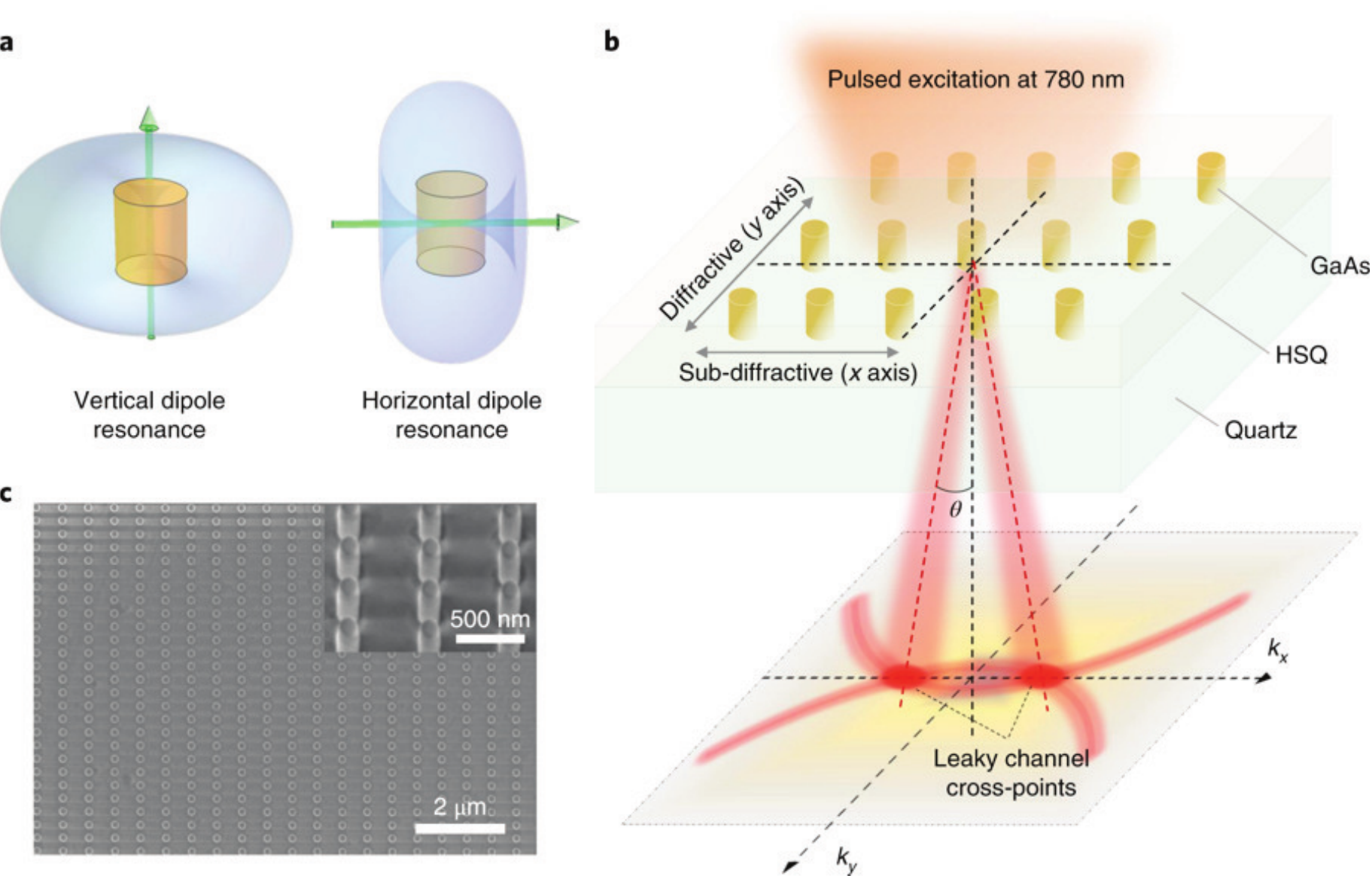}
  \caption{
      \label{figJ}
        \textbf{Unidirectional BIC laser.} (a) Resonances in GaAs nanoparticles. (b) Sketch of the nanoparticle lattice growth on a quartz substrate embedded in hydrogen silsesquioxane resist. (c) Scanning electron microscope (SEM) image of the metasurface.
       Adapted with permission from~\cite{Ha-2018-NN}.
	}
\end{figure*}

Following this idea, the authors proposed two metadevices: an optical router and an ultrafast optical pulse generator, both implemented on-chip. The router setup consisted of four silicon nanowires with a $100\times\SI{150}{nm}$ rectangular cross section arranged in a cross configuration and with the InGaAs nanodisk at the center, as shown in Fig.~\ref{fig6}.

Taking advantage of the asymmetry of the near-field of the excited anapole mode, the device  couples  the impinging light into the waveguide orthogonal to the polarization. The ultrafast pulse generator source, conversely, exploited the near-field synchronization mechanism of several anapoles close to each others.
The structure offers an alternative to other technologies such as $Q$-switching components and saturable absorbers, which require macroscopically large footprints.

This work showed that it is possible to generate a mode-locked signal in a nanowire, by assembling a line of several nanodisks next to a tapered structure, where each nanodisk is designed to sustain an anapole mode at different frequencies within the InGaAs emission band (different radii).
The coupling between the anapole can also be manipulated by shifting the position of each nanodisks, which can result in the generation of ultrashort pulses with different characteristics ($\SI{145}{fs}$ length with a $\SI{1.65}{ps}$ period in the example provided by the authors).
Numerical simulations performed in~\cite{Totero-2017-NC} showed that the anapole chain behave very closely to an optical neural network, opening also future applications of anapoles for optical integrated neurocomputing and neuromorphic photonics circuits.


Concerning the exploitation of BIC for lasing action, the first proposal was based on a photonic crystal membrane by Hirose and co-workers~\cite{Hirose-2014-NP},
though as noted above the authors did not report the presence of BIC. The recent work of Kodigala and co-workers~\cite{Kodigala-2017-N} proposed the first BIC laser designed for telecommunication wavelength ($\lambda=\SI{1.55}{\micro m}$) and relying on semiconductors. The cavity in this system is represented on Fig.~\ref{figH}. It is composed of an array of connected semiconductor nanodisks (with thickness of $\SI{300}{nm}$ and radii varying from 500 to $\SI{550}{nm}$) interconnected by bridges and arranged in a thin membrane suspended in air.
Through numerical simulations, the authors found an optimal radius of $\SI{528.4}{nm}$ for the $Q$ factor, which corresponds to the geometry for which two modes close to the excitation frequency create BIC states as they decouple from the radiation continuum. The lasing phenomenon was experimentally demonstrated while pumped by a pulsed laser ($\lambda=\SI{1064}{nm}$) on arrays of sizes ranging from $8\times8$ to $20\times20$ at room temperature, and show very good agreement with theoretical predictions.

The first BIC-based lasing cavity relying on an all-dielectric nanoparticle system and operating at visible light was proposed in~\cite{Ha-2018-NN}, with a cavity composed resonant GaAs vertical cylinders (with a height of $\SI{250}{nm}$ and radius $\SI{50}{nm}$) arranged into a metasurface (see Fig. \ref{figJ}).
The metasurface is designed to be sub-diffractive in one direction but diffractive in the other (with periods of $\SI{300}{nm}$ and $\SI{540}{nm}$ respectively).\\
This configuration created a radiation channel that would not exist in a completely sub-diffractive array, in which a perfect BIC with infinite $Q$ factor would be formed by the in phase oscillation of vertical dipoles. Instead, a leaky resonance with finite but high $Q$ factor was formed in this system, resulting in a leaky channel. Lasing occurs at the angle corresponding to the crossing point of the emission bands defined by this leaky channel.

Lasing wavelength and emission angle was demonstrated to be tunable by varying the geometrical parameters of the lattice and the peak of the gain spectrum of GaAs (through temperature). The temperature range of operation is in this work limited to 77--$\SI{200}{K}$, though the authors expect to achieve room temperature operation with a higher gain material or by improving the emission efficiency of GaAs.

\section*{Summary and outlook}

We have discussed two types of exotic localized states of light, {\em anapoles} and {\em bound states in the continuum}, which attracted a special attention recently, and provide different ways to achieve large enhancement of the electromagnetic fields and the confinement of light in high-index dielectric resonators with significant advantages for integrated photonics applications. Further exploitation of the physics of these nonradiating states and the study of their properties would involve a design of different types of dielectric resonant structures with a higher field enhancement and stronger magnetic field confinement in specific "hot-spots", improving the values of the $Q$ factor which were achieved previously only with the help of plasmonic structures, and also the study of nonlinear and quantum effects.  

Importantly, the use of high-index dielectric materials that replace noble metals in plasmonics present a particular advantage from both cost and scalability perspectives, especially considering that anapoles and BIC are generally engineered in rather simple geometrical structures that can be integrated into an on-chip circuitry. The use of resonant structures may improve substantially both performance and efficiency of subwavelength integrated photonics actively discussed in the current literature~\cite{nature_david,kita_optica}. 

As many effects with high-index dielectric resonators have now been understood and demonstrated, all the elements are now in place for the next step of assembling all-dielectric technological applications in different domain of nanophotonics, including, but not limited to, biological sensing, Raman scattering, multifunctional on-chip lasers, as well as integrated neurophotonics architectures.

It is also worth noticing that these studies stimulated the development of other approaches for achieving high efficiencies in all-dielectric resonant optical structures; as an example, We mention here a very recent work relying on the excitation of toroidal dipolar response (not exactly at the anapole frequency), and based on $E$-shaped all-dielectric metasurfaces operating in the infrared frequency range~\cite{Han-2018-OME}.

Harnessing of these novel nonradiating states in photonics has therefore a great potential to open a new era for all-dielectric metadevices in nanophotonics and meta-optics.

\section*{Acknowledgments} 

The authors acknowledge productive discussions and useful collaboration with many of their  colleagues, co-authors, and students. Especially, they are indebted to L.~Carletti, S.~Kruk, A.~Kuznetsov, M.~Limonov, B.~Lukyanchuk, M.~Rybin, D.~Smirnova, J.~Totero Gongora, and V.~Tuz.  They also acknowledge a financial support from KAUST (OSR-2016-CRG5-2995), the Australian Research Council, the Strategic Fund of the Australian National University, the Russian Foundation for Basic Research (18-32-20205), the Ministry of Education and Science of the Russian Federation (No.
3.1365.2017/4.6) and the Grant of the President of the Russian Federation (MK-403.2018.2). K.K. and A.B. acknowledge a support from the Foundation for the Advancement of Theoretical Physics and Mathematics "BASIS" (Russia).


\end{document}